\documentclass[twocolumn, twoside,prl]{revtex4-1}

\usepackage[english]{babel}

\usepackage{graphicx}

\begin{document}

\title{Phase estimation with weak measurement using a white light source. }

\author{Xiao-Ye Xu$^1$}

\author{Yaron Kedem$^2$}

\author{Kai Sun$^1$}

\author{Lev Vaidman$^2$}

\author{Chuan-Feng Li$^1$}

\author{Guang-Can Guo$^1$}

\affiliation{$^1$ Key Laboratory of Quantum Information, University of Science and Technology of China, CAS, Hefei, 230026, People's Republic of China}

\affiliation{$^2$ Raymond and Beverly Sackler School of Physics and Astronomy, Tel-Aviv University, Tel-Aviv 69978, Israel}

\begin{abstract}
We report results of a high precision phase estimation based  on  a weak measurements scheme using commercial light-emitting diode. The method is based on a measurement of the imaginary part of the weak value of a polarization operator. The imaginary part of the weak value appeared due to the measurement interaction itself. The sensitivity of our method is equivalent to resolving  light pulses of  order of  attosecond and it is robust against chromatic dispersion.
\end{abstract}

\maketitle

High precision phase measurements play a significant role in modern physics.  The standard tool is an interferometer with  a balanced homodyne detection \cite{Caves}.  It requires a coherent source and the precision is dominated by the intrinsic quantum noise \cite{Klauder}.  To reduce the influence of the noise, quantum metrology technologies \cite{metro} including N00N states \cite{noon} and squeezed states \cite{squeez} have been exploited, while white light is usually deemed to be useless in quantum metrology. Recently it has been proposed that  white light can  be used for a very precise phase estimation \cite{Brunner,Li}, when   weak measurements are performed. Here we experimentally demonstrate such a sensitive method utilizing white light from a commercial light-emitting diode (LED). This opens a new avenue for a high-resolution phase estimation.

As in other weak measurement experiments in which the  Aharonov-Vaidman-Albert (AAV) amplification effect \cite{aav} was demonstrated, we measure the photon polarization operator $A$  with eigenvalues 1 and -1 for the two orthogonal polarizations. The polarization can be pre- and post-selected with a very good precision. The role of the measuring device is played by the spatial degree of freedom of  light. In most weak measurement experiments, the relevant spatial degree of freedom is the position in the transverse direction, i.e. perpendicular to the direction of the light propagation.  Here we consider, instead, the longitudinal direction \cite{Brunner,Li}.

\begin{figure}
\centering
\includegraphics[width=0.5\textwidth]{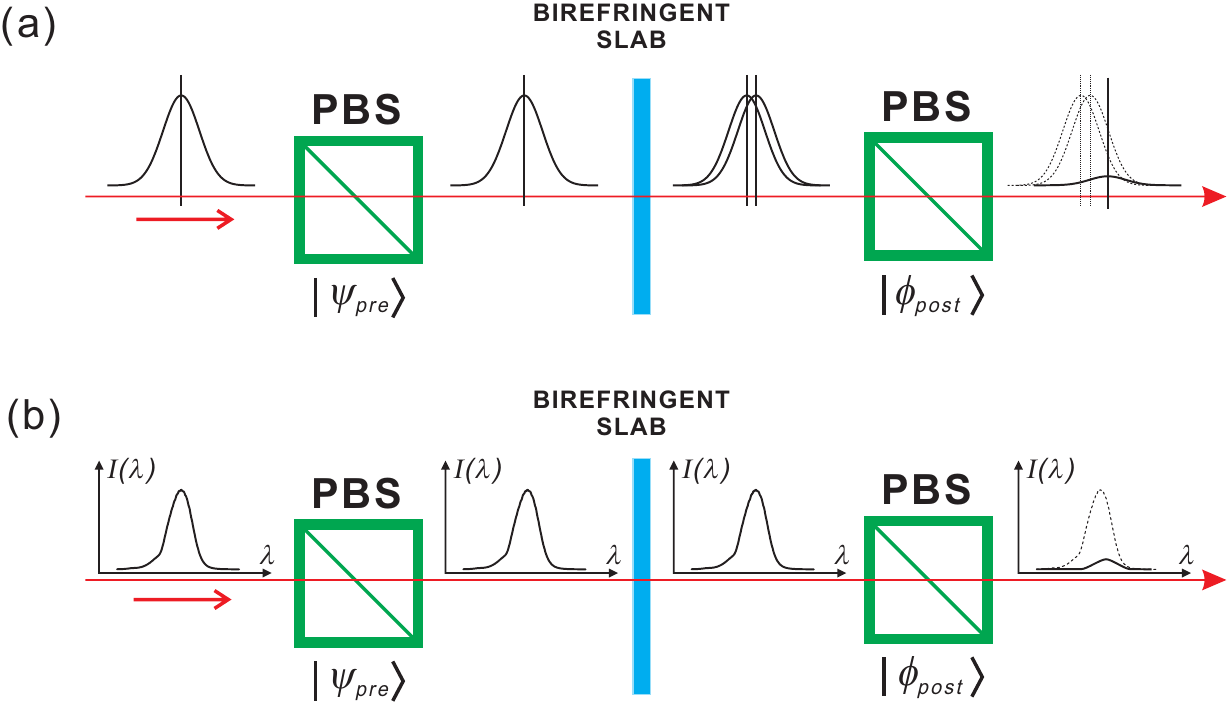}
\caption{ Weak measurement of the photon polarization.  Photons emitted from the source  are preselected by a  polarization beamsplitter (PBS) in a state $\left|\psi_{pre}\right\rangle$,  undergo weak measurement interaction by passing through birefringent plate and are post-selected at a nearly orthogonal state $\left|\phi_{post}\right\rangle$  by a second PBS.~ a). The wave packets with orthogonal polarizations are delayed after birefringent plate one relative to  the other  by a few attoseconds, but after the second PBS they interfere in the post-selected wave packet being shifted  in proportion to the real part of the weak value of the polarization operator. b). The shift of the spectrum of the light is proportional to the imaginary part  of the weak value of the polarization operator.
  \label{scheme}
}
\label{setup}
\end{figure}

The interaction Hamiltonian is
\begin{equation}\label{h}
H = g(t) PA ,
\end{equation}
where $g(t)$ is the coupling strength satisfying $\int g(t) dt = k $ and $P$ is a component of the momentum of the photon.
In the first realization of weak measurement the transversal shift was created by a tilted plate of a birefringent material \cite{Ritchie}. In our experiment, the plate is placed perpendicularly to the photon's velocity and leads to a longitudinal shift, see Fig. \ref{scheme}a. We consider a very thin birefringent  plate which leads to a time delay of  a  few attoseconds between the wave packets  with different polarizations. The AAV effect with the proper pre- and post-selection of polarization can increase the time delay significantly, but a truly dramatic advantage is obtained for  the  measurement of the imaginary part of the weak value of the polarization operator  which corresponds to the measurement of the spectrum shift, see Fig.  \ref{scheme}b.

In weak measurement experiments with the transversal shift, the component of momentum in the interaction Hamiltonian (\ref{h}) is perpendicular to the direction of light and its expectation value is 0. In the weak measurement regime, the uncertainty of the perpendicular momentum is small and thus the interaction Hamiltonian is weak and we can neglect the change of the photon polarization wave function due to the measurement interaction.

 In our case, the momentum in the Hamiltonian  (\ref{h}) is in the direction of the motion and its expectation value is $P_0$. Thus, the evolution of the photon's polarization during the measurement cannot be neglected. The uncertainty of the momentum $\Delta P$, corresponding to the width of the spectrum of our light source, is relatively small, so the evolution of the polarization state is essentially known and can be taken into account. Since the uncertain part of the interaction is small we can use the formalism of weak measurements even when the coupling to the measurement device causes a non-negligible effect.

An important practical advantage of weak measurements lies in measuring the imaginary part of the weak value, which  shifts  a variable conjugate to the one which is normally affected by the relevant Hamiltonian \cite{Joz}. This opens a whole new category of experimental techniques. Indeed, it led to the first observation of the tiny shift due to the quantum Hall effect for light \cite{HK} and to ultrasensitive measurement of the mirror tilt in a Sagnac interferometer \cite{howel}.

It has been suggested that the measurement of a phase using imaginary weak value  can outperform standard interferometry \cite{Brunner}.
However, in our experiment, the pre- and post-selection of polarization is achieved using linear polarizers, so in the absence of additional evolution, the weak value of a linear polarization operator is real. Inserting a quarter-wave plate will create a circular component in the polarization, but because of the wide spectrum, the polarization state will be a mixture with a large uncertainty and for such a wide spectrum we cannot correct it using a dispersive material. Instead, the imaginary part of the weak value is created by the measurement interaction due to the non-negligible evolution  mentioned above.

When  light of wavelength $\lambda_0$ (corresponding to momentum $P_0$) goes through a half wave plate of width  $L_0 $, the polarization state, represented by a vector on the Poincare sphere, is rotated by the angle of $\pi$. For a plate of width $L$ the rotation angle will be $\alpha = \pi { L \over L_0 }$. This corresponds to a strength of the interaction (\ref{h}) of $k = \alpha / P_0$.

The polarizations states that are created and postselected by the prisms are
\begin{eqnarray}
\left|\psi_{pre}\right\rangle & = & \sin {\pi \over 4}~  \vert H\rangle+ \cos {\pi \over 4}~ \vert V\rangle   , \\
\left|\phi_{post}\right\rangle & = & \sin\left({\beta \over 2} - { \pi \over 4} \right)\vert H\rangle+\cos\left({\beta \over 2} - { \pi \over 4}\right)\vert V\rangle,
\end{eqnarray}
where $ { \pi \over 2} - {\beta \over 2}$ is the angle between the optical axis of the prisms so $\pi - \beta$ is the angel between the representations of $\left|\psi_{pre}\right\rangle$ and $\left|\phi_{post}\right\rangle$  on the Poincare sphere.

Consider the polarization state of a photon at a position $x$ in between the boundaries of our effective thin plate. It is separated by a plate of width $x$ from the preparing prism and a  plate of width $L-x$ from the post-selecting prism. The two-state vector $\langle \phi |~|\psi\rangle $ of polarization at point $x$ is
\begin{eqnarray}
\left|\psi\right\rangle & =
 & {1 \over \sqrt{2}} \left( e^{-i {\pi x  \over 2 L_0 }} \vert H\rangle+ e^{i{\pi x  \over 2 L_0 }} \vert V \rangle \right) \nonumber \\
\left\langle \phi \right| & = & e^{-i{\pi (L-x)  \over 2 L_0 }} \sin\left({\beta \over 2} - { \pi \over 4}\right) \left\langle H \right| \nonumber \\
&+&  e^{i{\pi (L-x)  \over 2 L_0 }} \cos\left({\beta \over 2} - { \pi \over 4}\right)\left\langle V \right|.
\end{eqnarray}
The weak value of $A=\vert H\rangle\left\langle H\right| - \left|V\right\rangle \left\langle V\right|$ turns out to be the same for any point $x$:
\begin{eqnarray}
A_w\equiv\frac{\left\langle \phi\right|A\left|\psi\right\rangle }{\left\langle \phi|\psi\right\rangle } & = &
\frac{ 1 - e^{i \alpha} \cot\left({\beta \over 2} - { \pi \over 4}\right)}{1 + e^{i \alpha} \cot\left({\beta \over 2} - { \pi \over 4}\right)}.
\end{eqnarray}

For small tilt $\alpha \ll1,$ and almost orthogonal post-selected state $\beta\ll1$, we obtain:
\begin{equation}
A_{w} \simeq \frac{1}{ \beta-i \alpha},
\end{equation}
and
\begin{equation}\label{IwvS}
{\text Im}A_{w}  \simeq \frac{\alpha}{\beta^{2}+\alpha^{2}}.
\end{equation}

When the conditions of the weak measurements are fulfilled, the shift in the expectation value of $P$, after the interaction (\ref{h}), is given by $\delta P = 2 k (\Delta P)^2 {\rm Im} A_w $ \cite{AV90}.   For our setup, the coupling,  $k = \alpha / P_0$, is very small. The longitudinal shift is much smaller than  the wavelength,  so we have:
\begin{equation} \label{P}
\delta P = \frac{2 \Delta P^2 \alpha^2}{ P_0 \left( \beta^{2}+\alpha^{2} \right)}.
\end{equation}

Since $\Delta P \ll P_0$ and the light has just one direction, we can measure $P$ using a spectrometer, based on the relation $\lambda = {2 \pi \over P} $. Then, instead of equation (\ref{P}) we will have an equation for the shift in the wavelength:
\begin{equation} \label{shiftL}
 \delta \lambda = \frac{2  \Delta \lambda^2 \alpha^2}{ \lambda_0 \left( \beta^{2}+\alpha^{2} \right)}.
\end{equation}

We can see here the advantage of using white light with a wide spectrum. Our measured signal is $\delta \lambda$ and it is proportional to $\Delta \lambda^2$, so the signal to noise ratio is proportional to the uncertainty $\Delta \lambda$.
\begin{figure}
\centering
\includegraphics[width=0.5\textwidth]{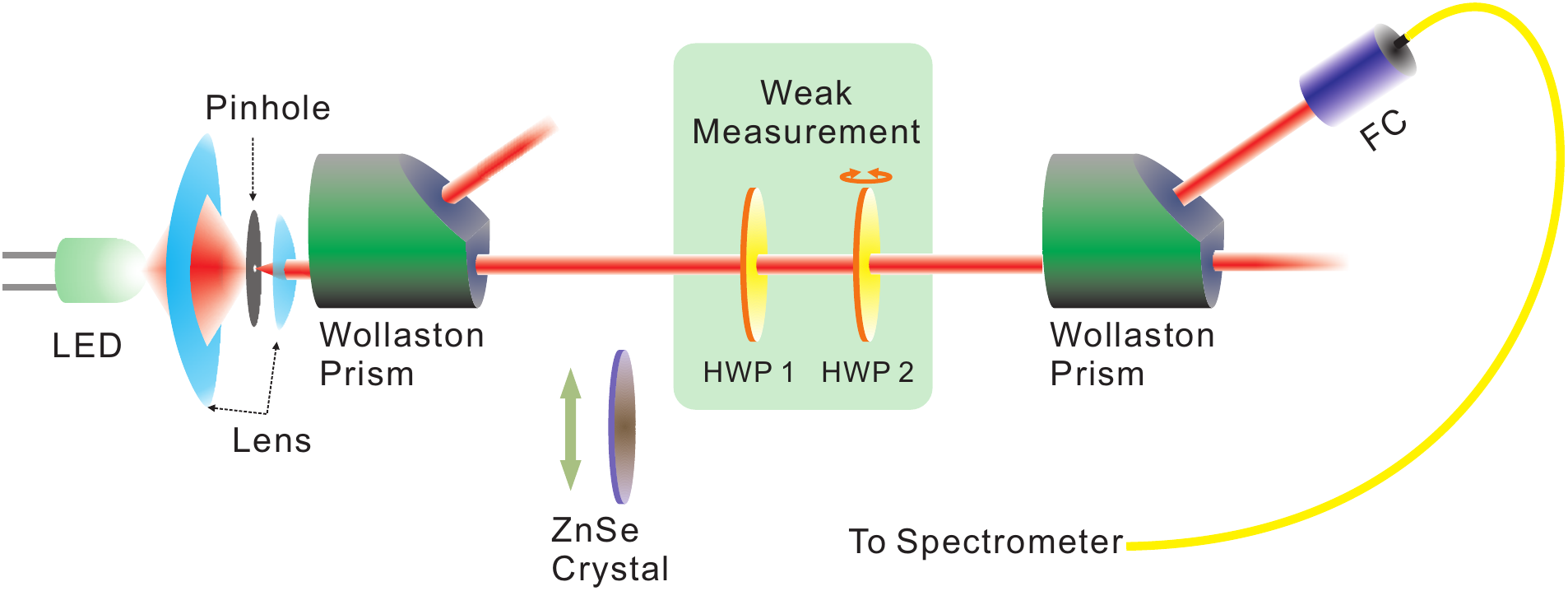}
\caption{ Detailed experimental setup: Photons emitted from a LED (produced by Epitex company), with measured central wavelength $\lambda_0 =  808 nm$ and spectral width of $\Delta\lambda = 38.8nm$ were collected by two lenses, pass through a pinhole and enter the first PBS (a Wollaston prism) which preselects the polarization state $\left|\psi_{pre}\right\rangle$. Then the light passes two true zero order half wave plates (HWPs) with their optical axes perpendicular to each other and at $45^{\circ}$ to the axis of the first prism. The plane of the second HWP is tilted by the angle $\theta$ with respect to the first HWP. A second Wollaston prism with its optical axis at an angle ${\pi\over 2} - {\beta \over 2}$ with respect to the first one postselects  a polarization state which is nearly orthogonal to the preselected one. The light is then collected by a single-strand optical fiber and sent to spectrometer with sampling period of $0.02nm$ and range $715-915nm$. A $1mm$ thick ZnSe crystal is inserted to introduce a strong dispersion. FC - fiber collector.
}
\label{setup}
\end{figure}

The weak measurement interaction in our experiment  is between the polarization and the spatial degrees of freedom of photons. Conceptually, such interaction is achieved by placing a plate of birefringent material perpendicularly to the photon velocity. The effect has to be very small and in practice, we use, instead, two identical true zero order half-wave plates (HWP), one perpendicular and one almost perpendicular but with a tiny tilt, see Fig. \ref{setup}. The optical axes of the two HWPs are perpendicular to each other such that their effects cancel each other and the total longitudinal relative shift of the different polarizations vanishes when the tilt is zero. A tilt increases the optical path in the tilted HWP so that the system of the two HWPs becomes equivalent to a plate of a very small width of the same material and orientation as the tilted one. The reason for using such a construction of the two HWPs is the practical difficulty of constructing and manipulating a very thin birefringent plate and simplicity of changing the (effective) width of the plate by tilting. The correspondence between the phase shift $\alpha$ and the tilt angle $\theta$, shown at the lower part of Fig. 3.,  is
\begin{eqnarray}\label{L}
\alpha = \pi { L \over L_0 } =\pi \left({1 \over \sqrt{1- \sin^2{\theta} / n_0^2}}-1\right)  \simeq { \pi \theta^2 \over 2 n_0^2},
\end{eqnarray}
where refractive index $n_0 =1.54$.
\begin{figure}
\centering
\includegraphics[width=0.5\textwidth]{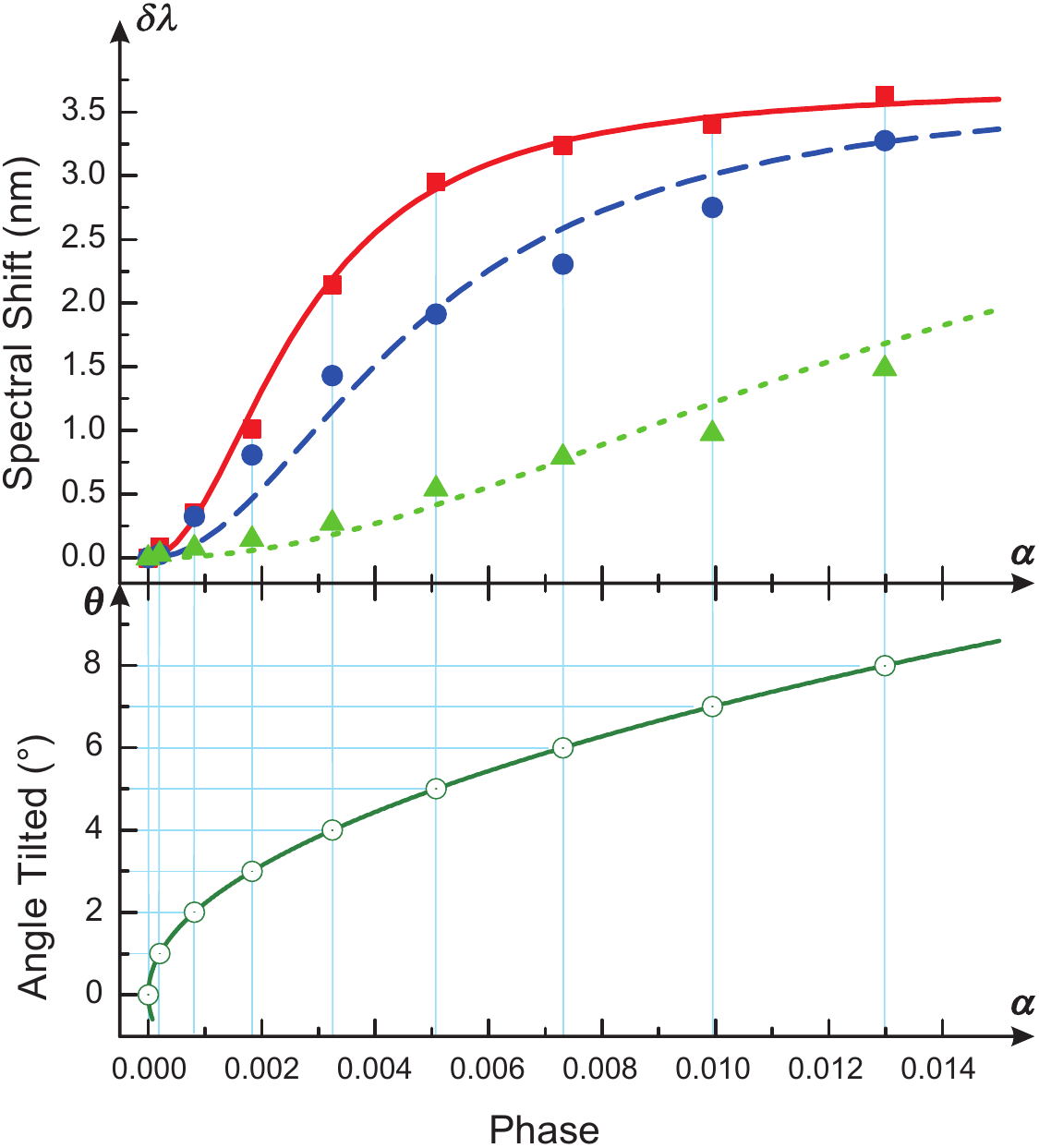}
\caption{Experimental results. The  shift of the  spectrum of light $\delta \lambda$ as a function of a corresponding phase shift $\alpha $.
The spectral shift are shown for three different values of post-selection parameter $ \beta$. The red-square, blue-circle and green-triangle points correspond to $ \beta = 0.000, 0.004$ and $0.014$ respectively. The red, blue and green lines are the theoretical predictions. The lower part shows the connection between  $\alpha $ and the tilt angle $\theta$ which is controlled in the experiment as explained in the text, Eq. \ref{L}.
}
\label{result}
\end{figure}

The theoretical formula (\ref{shiftL}) assumed ideal polarizers. However, for almost orthogonal post-selection corresponding to a very small parameter $\beta$, we have to take into account the uncertainty in the polarization of the photon passing through the polarizers. When we set the angle between the optical axis of the prisms
 $ { \pi \over 2} - {\beta \over 2}$, the probability for the actual parameter $\beta'$ of the  pre- and post-selected photon is proportional to $\beta'^2 e^{\frac{(\beta-\beta')2}{2\Delta^2}}$. Factor $\beta'^2$ is an approximation due to the probability of the post-selection $cos^2{\frac{\beta'}{2}}$. In our experiment $\Delta \sim 0.0027$.  Due to this uncertainty,  setting the polarisers orthogonal to each other, $\beta=0$,  leads to the effect similar to a setup with ideal polarizers set to a parameter $\beta \sim 0.002$.  This happens because our effect is not sensitive to the sign of $\beta$.

The  experimental results, along with theoretical curves obtained by averaging of (\ref{shiftL}) corresponding to the polarization uncertainty are shown in Fig. \ref{result}. We have performed measurements for three post-selected states with $\beta = 0.000, 0.004$ and $0.014 $. In all cases we changed  $\alpha$ from 0  to $0.013$ by changing the effective width of the birefringent plate. We obtained good correspondence between theory and experiment and especially good correspondence for orthogonal polarization filters, the case which was easiest to control.

In order to test the robustness of our method to chromatic dispersion, we introduced it artificially using a 1mm thick ZnSe crystal, see Fig. \ref{setup}. A few femtosecond pulse of light going through the crystal will be broadened by order of hundreds femtosecond, an effect that can gravely harm the precision for many setups. Our method however, worked well also when the ZnSe crystal was moved inside.  The crystal variable spectrum transmittance slightly changed the spectrum: $\lambda_0 =  805nm$, $\Delta \lambda =41.6nm$.  The results for $\beta = 0.000, 0.004$ are presented in Fig. 4.  The change relative to the case without  ZnSe crytsal is small. This demonstrates an important advantage of our method. Modern metrology technologies methods and the scheme of the measurement of the real part of the weak value of the polarization requires coherent source. The amplification in the measurement of the imaginary part of the weak value works also for our white light source.

On Fig. 4 we also presented our test of the theory in which we  repeated the experiment for $\beta =0.004$ with the crystal filtering the LED light and thus reducing the spectrum width to $\Delta \lambda =18.9nm$, $\lambda_0 =  795nm$.  Reduction of the shift of the spectrum demonstrates the advantage of the wide spectrum of our LED source.

\begin{figure}
\centering
\includegraphics[width=0.5\textwidth]{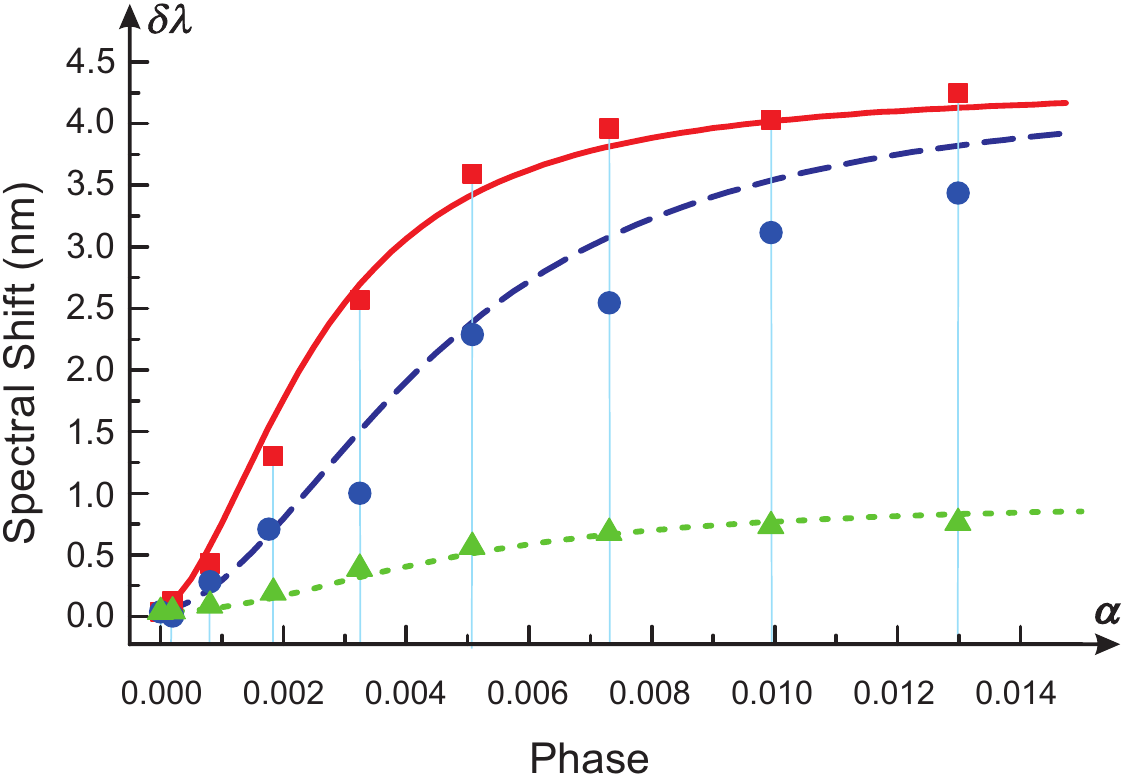}
\caption{The  shift of the  spectrum of light $\delta \lambda$ for light passing through thick ZnSe crystal.  The red-square, blue-circle  points correspond to  $\beta = 0.000$ and $ 0.004$  respectively. The green-triangles also correspond to  $ \beta = 0.004$, but for filtered light having significantly smaller spectrum width. The red, blue and green lines are the theoretical predictions.
}
\label{result2}
\end{figure}

Since our expression (\ref{shiftL}) for the shift is not just linear in the phase, a special analysis is needed in order to estimate the precision of our method. In order to get the best precision we have to tune the post-selection parameter $\beta$ depending on the value of the measured phase $\alpha$. Taking partial derivative of (3) relative to $\alpha$, we find that the optimal precision is obtained when we choose $\beta \simeq \alpha$. For this choice of $\beta$, the uncertainty is $ \Delta \alpha ={\lambda_0 ~\alpha \over \Delta \lambda ^2 }\Delta(\delta \lambda)$. In our experiment, estimating the measured uncertainty of the wavelength as $\Delta(\delta \lambda)=0.1nm$, we obtain $ \Delta \alpha \simeq 0.1\alpha$. For  very small phases, we set $\beta=0$ and utilize the uncertainty of the polarizers, so (\ref{shiftL}) is irrelevant. In this case, the simplest way to estimate the precision is by viewing the theoretical curve as a calibration in light of the very good correspondence of $\beta=0$ curve with the experimental results. We see that the phase shift  $\alpha \simeq 10^{-3} $ can be estimated with precision of the order of $\alpha \simeq 10^{-4} $. Our results compete well with coherent light phase weak measurements \cite{Starling}  and currently are significantly better than quantum metrology technologies measurements using  N00N  and squeezed states \cite{sub}, which are still in the process of solving experimental problems \cite{real}.

We have performed an experiment using commercial LED source demonstrating a new method for precision phase estimation based on weak measurement. The method  is invulnerable to chromatic dispersion. Its simplicity and robustness suggest that it can be applied for a vast range of applications.

After completion of our experiment,  we have learned about another theoretical modification  \cite{Struebi} which includes analysis of the spectrum of transmitted and reflected light of the post-selection polarization beam splitter. It might lead to further improvement of the precision of the phase estimation, but it is beyond the scope of the current work.

 This work has been supported in part by the National Basic Research Program of China (Grant No. 2011CB921200), National Natural Science Foundation of China (Grant Nos. 60921091, 11274289), and  the Israel Science Foundation  Grant No. 1125/10.

\end{document}